\documentclass[12pt]{article}

\usepackage{epsfig}
\usepackage{ams-mdbch}

\begin{document}

\begin{center}

{\Large EVOLUTIONARY CONSTRAINTS:}

\vspace{8mm}

{\Large GAUSS' LAW AS A TOY MODEL FOR GLUING}

\vspace{10mm}

{\large Ingemar Bengtsson$^*$}

\vspace{6mm}

{\large Istv{\'a}n R{\'a}cz$^{\dagger}$}

\vspace{10mm}

{\small 
${}^{*}${\sl Stockholms Universitet, AlbaNova,}

{\sl SE-106 91 Stockholm, Sverige}

\

${}^{\dagger}${\sl HUN-REN Wigner RCP, H-1121 Budapest,}

{\sl Konkoly-Thege Mikl{\'o}s {\'u}t 29-33, Hungary}
}

\vspace{10mm} 

{\bf Abstract:}

\end{center}

{\small 
\noindent It is possible to solve the Einstein constraint equations as an 
evolutionary rather than an elliptic system. Here we consider the Gauss constraint 
in electrodynamics as a toy model for this. We use a combination of the evolutionary method  with the gluing  construction to produce initial data for an electromagnetic pulse  surrounded by vacuum. It turns out that solving the evolutionary form of the constraint is straightforward, and explicitly yields the desired type of initial data. In contrast, proving the existence of a solution to the same problem within the 
elliptic setting requires sophisticated arguments based on functional analysis.
}

\vspace{12mm}

{\bf 1. Introduction}

\

\noindent Space and time are fundamental concepts in science, especially in the 
various branches of physics. However, no other theory has touched the core of these concepts more than relativity. It is fair to say that the revision of the concept of time was the key step that led to the birth of special and general relativity as theories. In these theories we learn that time behaves in many ways like space. This in turn led to concepts like spacetime and the metric theory of gravity. We also learn from Penrose's cosmic censor hypothesis that generic spacetimes should be maximal Cauchy expansions, which means that the theory is ``predictable'' \cite{P}. On a technical level, this requires some kind of separation of space and time in order to make sense of statements like ``spatial geometry evolves in time''. This happens despite the fact that there is no 
intrinsic way of dividing spacetime into space and time. 
There are, of course, various attempts to make meaningful distinctions. Based on one of Geroch's results \cite{G}, one could boldly claim that global hyperbolicity ensures that the spacetime manifold M, which is considered to be the space of predictable events, is topologically equivalent to a product 
$\Sigma \times \mathbb{R}$,  
where $\Sigma$ is a spatial Cauchy surface in M, while time should correspond to the one-dimensional factor of the above product. Based on analogous views, a well-known philosopher of science was led to say that ``a temporal direction at a point in spacetime is a direction in which our best theory tells the strongest, i.e. most informative, story'' \cite{C}. 
The light-cone structure in spacetime, that we use to distinguish spatial 
and temporal directions, is closely related to the symmetrizable hyperbolic character of our fundamental field equations. In most cases, the characteristic cones associated with these field equations, which are determined by the principal part of the PDEs, coincide with that of the null cone of the spacetime metric. This is the case for Einstein's field equations, Maxwell's equations, the Klein-Gordon equation, and many others.

Close to the centenary of general relativity, new ideas about the basic concepts of space and time have emerged. It was found that time evolution can occur in Einsteinian spaces with Riemannian metrics, where in principle there is no place for the concept of time \cite{BIH,Racz1,Racz3}. It is worth noting that in classical 
physics those equations which are formulated on spaces with positive definite 
metric are systematically formulated as elliptic partial differential equations 
and solved accordingly. The new results discussed in \cite{BIH,Racz1,Racz3} challenge 
this traditional view. 
The main message is that any spatial coordinate, or any real function of spatial coordinates, whose level sets are diffeomorphic to a 
manifold $\sigma$, can also behave like time, provided that space has the product topology 
$\sigma \times \mathbb{R}$,
Based on these observations, it was shown by one of us \cite{Racz1} 
that the constraint equations (for some suitably chosen constrained 
variables) can be solved as either parabolic-hyperbolic or algebraic-hyperbolic 
systems. The idea has already scored some successes when applied 
to general relativity \cite{Racz2, Beyer, Beyer2, Racz4}, but we feel that it 
would be of interest to have a simpler example available. 

The present work is intended to demonstrate the capabilities of these new ideas when 
applied to the constraint equation in electrodynamics. To keep the discussion 
fully concrete we choose a concrete question: 
How do we find initial data for a sourceless electric field confined inside 
a ball of radius $R_1$, and vanishing outside some concentric ball with radius 
$R_2 > R_1$? It should be possible to produce a 
fair approximation of such a solution by means of a pulse from a 
free-electron laser, so we expect the solution to exist. The question is of 
some interest in itself, but it has also been raised as a toy example of the gluing constructions that 
have been made for initial data of Einstein's equations. Using 
the elliptic method one is then led to a highly technical proof of the existence 
of a solution \cite{Chrusciel, Corvino, Delay}. The point we want 
to make is that the evolutionary method trivializes the existence proof, 
and can produce solutions in explicit form. 
 
Hence we are concerned with smooth solutions to Gauss' law  

\begin{equation} \nabla \cdot {\bf E} = 0 \end{equation}

\noindent in flat space. This is an underdetermined partial differential equation 
for the electric field, and we will treat it as an evolution equation instead 
of as elliptic equation. The idea is to prescribe the value of the 
field on some sphere in space, and then integrate outwards (or inwards) from 
there. Thus the evolution takes place in space, not in time; the outcome of 
this evolution is a set of initial data for Maxwell's equations.  For Gauss' law we give the 
general solution to our problem in Section 2, and we give a simple explicit 
example in Section 3. 

A comment on the evolutionary approach to the Einstein constraints in general 
relativity is in order here. Physically useful initial data should be asymptotically flat in a 
suitable sense. From our point of view this means that one is imposing conditions on two 
`time' slices when solving for the evolution in space. This kind of problem is not at all 
well understood, and has so far been discussed mostly in the context of stochastic evolution 
\cite{Schulman}. However, it is helpful that the equations we consider are underdetermined, 
and that a large part of the solution will consist of freely specifiable data. In the example 
we consider the analogous problem can be dealt with because we are able to work with a general 
exact solution. For the Einstein constraints the work that has been done so far gives good 
reasons to believe that the problem can be handled there as well: Asymptotically flat initial data 
have been obtained using the evolutionary approach  \cite{Racz2, Beyer, Beyer2, Racz4}.


\vspace{8mm}

{\bf 2. Solving the Gauss constraint}

\

\noindent To begin with, and to get some ideas for the freely specifiable data 
that we need, we will sketch the elliptic treatment of ref. \cite{Chrusciel}. 
We introduce two smooth 
functions $\chi = \chi (r)$ and $\psi = \psi (r)$ such that $\chi$ equals 1 when 
$r < R_1$, $\chi = 0$ when $r> R_2$, while $\psi$ is non-zero only inside the 
interval $r \in [R_1,R_2]$. Smooth functions having these 
properties are easy to find. Let us also introduce two divergence free 
vector fields ${\bf E}^{(1)}$ and ${\bf E}^{(2)}$, and a function $u$. 
Then we make the Ansatz 

\begin{equation} {\bf E} = \chi {\bf E}^{(1)} + (1 - \chi ){\bf E}^{(2)} 
+ \psi^2 \nabla u \ . \label{ansats} \end{equation}

\noindent This agrees with ${\bf E}^{(1)}$ inside the small ball and it 
agrees with ${\bf E}^{(2)}$ outside the larger ball, and setting its 
divergence to zero yields 

\begin{equation} \nabla \cdot {\bf E} = \nabla \chi \cdot ({\bf E}^{(1)} - 
{\bf E}^{(2)}) + \nabla \cdot (\psi^2\nabla u) = 0 \ . \label{Piotr} 
\end{equation}

\noindent Our problem is solved if we can solve this equation for the 
function $u$. In the elliptic setting the equation is addressed as it 
stands. The necessary existence theorem can be proved by constructing a 
functional whose would-be minima are solutions of this equation. It then 
requires some careful functional analysis to find a function space in 
which these minima actually exist \cite{Chrusciel, Corvino, Delay}. The 
point we are going to make is that the evolutionary formulation of Gauss' law 
simplifies the analysis very much. It also leads directly to an actual 
construction of the field. 

Using the evolutionary method we begin by 
foliating the region between the spheres with round spheres of radius $r$, 
having normal vectors $n_i = x_i/r$. Then we decompose our 
vector fields into normal and tangential parts, 

\begin{equation} E_i = \alpha n_i + L_i \ , \hspace{8mm} \alpha = 
n^iE_i \ . \end{equation} 

\noindent A calculation valid for a general foliation \cite{Racz3} shows that 

\begin{equation} \nabla \cdot {\bf E} = {\cal L}_{\vec{n}}\alpha + \alpha K + 
\bar{\nabla}_iL^i - \dot{n}^iL_i \  . \end{equation}

\noindent Here ${\cal L}_{\vec{n}}$ denotes the Lie derivative in the 
direction of the normal vector---the direction in which the informative 
story is going to be told. The divergence $\bar{\nabla}_iL^i$ uses the projected covariant 
derivative and can be written as $\bar{\nabla}_iL^i = \gamma^{ij}\nabla_iL_j$, where $\gamma_{ij}$ is the 
induced metric on the leaves. The mean curvature 
of the leaves is denoted by $K$, and we have defined $\dot{n}^i = n^j\nabla_jn^i$. In our 
case the leaves are round spheres with $K = 2/r$, the `time' function is $r$, $\dot{n}_i$ vanishes, 
and the equation simplifies to 

\begin{equation} \nabla \cdot {\bf E} = 
\partial_r\alpha + \frac{2}{r}\alpha + \bar{\nabla}_iL^i 
= \frac{1}{r^2}\partial_r(r^2\alpha) + \bar{\nabla}_iL^i \ . \label{eveq} \end{equation}

\noindent This is to be set equal to zero. The resulting equation is an 
evolution equation for $\alpha$, while the tangential part $L_i$ is to be 
regarded as freely specifiable data. 

The question that arises is how to choose $L_i$, the part of the field 
that is tangential to the leaves of foliation. Here the answer comes 
from the Ansatz (\ref{ansats}). The fields ${\bf E}^{(1)}$ and ${\bf E}^{(2)}$ are 
assumed to be known, and it is understood that they are divergence free, so we 
can write 

\begin{equation} \bar{\nabla}^iL_i^{(1)} = - \frac{1}{r^2}\partial_r(r^2\alpha^{(1)}) 
\end{equation}

\noindent for some known function $\alpha^{(1)}$, and similarly for ${\bf E}^{(2)}$. 
The function $\chi$ is a function of $r$ only, hence $\bar{\nabla}_i\chi = 0$, and 
similarly for $\psi$. With this understanding the evolution equation (\ref{eveq}), 
to be solved for the normal part $\alpha$, is 

\begin{eqnarray} \frac{1}{r^2}\partial_r (r^2\alpha ) = - \bar{\nabla}_iL^i = 
- \chi \bar{\nabla}^iL^{(1)}_i - (1-\chi )\bar{\nabla}^iL_i^{(2)} - \psi^2 
\bar{\nabla}_i\bar{\nabla}^iu = \nonumber \\ \\ = 
\frac{\chi}{r^2}\partial_r(r^2\alpha^{(1)}) + \frac{1-\chi}{r^2}\partial_r(r^2\alpha^{(2)}) 
- \frac{\psi^2}{r^2}\triangle_Su \ , \nonumber \end{eqnarray}

\noindent where $\triangle_S$ is the familiar Laplacian on the unit sphere and the 
function $u$ remains to be choosen. Our 
initial condition is that the field shall agree with ${\bf E}^{(1)}$ when $r = R_1$, 
so we set 

\begin{equation} \alpha_{|_{R_1}} = \alpha^{(1)}_{|_{R_1}} \ . \label{initial} 
\end{equation}

\noindent As mentioned above the drawback of the evolutionary formulation is that we have 
no {\it a priori} control of the asymptotics. In the present case this translates 
into the problem of how to ensure that the field agrees with ${\bf E}^{(2)}$ when 
$r = R_2$. There is still hope, because the freely specifiable data have not 
yet been fully specified. To achieve the aim we have to choose the function 
$u = u(\theta , \phi;r)$ in a suitable way. 

We begin by integrating our equation, which in this case can be done exactly: 

\begin{equation} r^2\alpha = \chi r^2\alpha^{(1)} + (1-\chi )r^2\alpha^{(2)} - 
\int_{R_1}^r\left( r^2(\alpha^{(1)} - \alpha^{(2)})\partial_r\chi + \psi^2 
\triangle_Su\right) {\rm d}r \ . \end{equation}

\noindent With this definition of $\alpha$ the electric field ${\bf E}$ 
obeys Gauss' law, and because of the way $\chi$ and $\psi$ were defined the 
initial condition that ${\bf E} = {\bf E}^{(1)}$ at $r = R_1$ 
holds. We also find 

\begin{equation} R_2^2\alpha_{|_{R_2}} = R_2^2\alpha^{(2)}_{|_{R_2}} - \int_{R_1}^{R_2} 
\left( r^2(\alpha^{(1)} - \alpha^{(2)})\partial_r\chi + \psi^2\triangle_Su 
\right) {\rm d}r \ . \end{equation}

\noindent Our problem will be solved if we can choose the function $u = u(\theta, 
\phi ; r)$ so that the integral vanishes. This task is much easier than that of 
addressing equation (\ref{Piotr}) directly. 

Let us make the problem more definite by insisting that ${\bf E}^{(2)} = {\bf 0}$, 
and let us also choose the smoothing functions so that 

\begin{equation} \psi^2(r) = - \partial_r\chi (r) \ . \label{smoother} \end{equation}

\noindent  Clearly $\psi$, so defined, has the properties that we demand. 
Then we require that 

\begin{equation} \int_{R_1}^{R_2} \psi^2 \left( r^2 \alpha^{(1)} - \triangle_S 
u\right) {\rm d}r = 0 \ . \label{integral} \end{equation}

\noindent If ${\bf E}^{(1)}$ is a Coulomb field then $\alpha^{(1)} = n_iE_i^{(1)} = 
q/r^2$, and we find a contradiction when integrating the equation over the 
unit sphere, because the integral of the Laplacian acting on a regular function 
of the sphere vanishes for each value of $r$. This was to be expected; by 
construction our vector field is divergence free in the shell between the spheres, 
and hence 

\begin{equation} \int_{r=R_1} {\bf E}\cdot {\bf da} = \int_{r=R_2} {\bf E}
\cdot {\bf da} \ . \end{equation}

\noindent More generally, if there is any non-zero charge distribution inside 
the ball then ${\bf E}$ cannot vanish 
at $r = R_2$. But this is the only restriction. If the integral of $\alpha^{(1)}$ 
over a sphere at fixed $r$ vanishes, as it must if ${\bf E}^{(2)}$ is sourceless, 
then it can be expanded using spherical harmonics with $\ell \neq 0$. It is then 
straightforward to give a similar expansion for a function $u$ such that 

\begin{equation} r^2 \alpha^{(1)} - \triangle_S u = 0 \ . \end{equation}

\noindent This is stronger than necessary for equation (\ref{integral}) to hold, but 
it shows without further ado that a solution exists. This is a great simplification 
compared to the existence proof offered in the elliptic setting \cite{Chrusciel}. 
Moreover the evolutionary existence proof is constructive, while the elliptic 
one is not. 

\vspace{8mm}

{\bf 3. An example}

\

\noindent An advantage provided by the evolutionary formulation of the 
constraint is that we can get explicit about the solutions. As a (very) 
simple example, let us consider a ball with radius $R_1 = 1$ containing constant 
crossed electric and magnetic fields, and let the fields vanish outside a 
ball with radius $R_2 = 2$. Say that 

\begin{equation} E_x^{(1)} = B_y^{(1)} = a \ , \end{equation}

\noindent all other components vanishing. With ${\bf E}^{(2)}$ vanishing we have 

\begin{equation} E_i = \alpha n_i + L_i = \alpha n_i + 
\chi L_i^{(1)} + \psi^2\bar{\nabla}_iu \ . \end{equation}

\noindent The tangential part consists of free data, and equation (\ref{smoother}) 
is in place. We choose the function $u$ to obey 

\begin{equation} \triangle_Su = r^2\alpha^{(1)} = axr = ar^2\cos{\phi}\sin{\theta} 
\ . \end{equation}

\noindent The solution is

\begin{equation} u = f(r) - \frac{a}{2}r^2\cos{\phi}\sin{\theta} = 
f(r) - \frac{a}{2}xr \ . \end{equation}

\noindent The function $f = f(r)$ need not concern us because only the 
tangential derivative of $u$ matters. When this choice is made our solution for 
the normal part of the field reads 
$\alpha = \chi \alpha^{(1)}$. It follows that 

\begin{equation} E_i = \chi \alpha^{(1)} n_i + \chi L_i^{(1)} + 
\psi^2\bar{\nabla}_iu= \chi E^{(1)}_i + \psi^2\bar{\nabla}_iu \ . \end{equation}

\noindent Explicitly, in Cartesian coordinates the solution is 

\begin{equation} {\bf E} = \frac{a}{2r}\left( \begin{array}{c} 2r\chi - 
(y^2+z^2)\psi^2 \\ xy\psi^2 \\ xz\psi^2 \end{array} \right) 
\hspace{8mm} {\bf B} = \frac{a}{2r}\left( \begin{array}{c} xy\psi^2 \\ 
2r\chi - (x^2+z^2)\psi^2 \\ yz\psi^2 \end{array} \right) 
\end{equation}

\noindent The magnetic field was similarly derived. 

\begin{figure}
        \centerline{ \hbox{
                \epsfig{figure=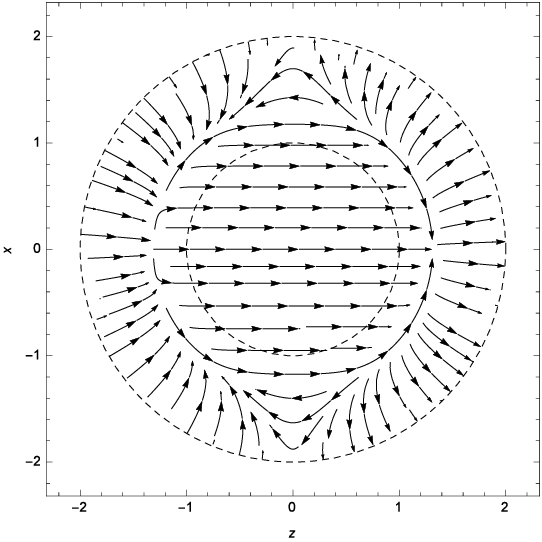,width=55mm}}}
        \caption{{\small The Poynting vector field in a simple solution. 
        The pulse is clearly about to move rightwards (in the $z$-direction).}}
        \label{fig:poynting}
\end{figure}

To get a hint about the time evolution of this blob of radiation 
we consider the Poynting vector field in Figure \ref{fig:poynting}. For 
illustrative purposes we have chosen 

\begin{equation} \psi = \left\{ \begin{array}{lcl} (r-1)^2(2-r)^2 & \mbox{if} 
& 1 < r < 2 \\ \\ 0 & \mbox{otherwise} & . \end{array} \right. \end{equation}

\noindent This implies finite differentiability only, but the solution 
can easily be made smooth with some slight modification. 

More interesting solutions, where the electric and magnetic fields are 
non-constant within the inner sphere, can be obtained if we solve the 
two-dimensional Poisson equation for these cases. 
However, since our purpose here 
was to highlight the advantages of the evolutionary formulation of the 
constraints, rather than to study pulses of electromagnetic radiation as 
such, we stop here. We believe that we have already made the point that 
the evolutionary formulation of the constraint equation can lead to 
a very informative story. 

\vspace{12mm}

\noindent \underline{Acknowledgement}: This project was supported in part 
by Hungarian Scientific Research fund NKFIH K-142423.

\

\end{document}